\documentclass[twocolumn,prl,10pt,amsmath,amssymb,nofootinbib,showpacs,superscriptaddress,floatfix]{revtex4-1}

\DeclareFontFamily{U}{rcjhbltx}{}
\DeclareFontShape{U}{rcjhbltx}{m}{n}{<->rcjhbltx}{}
\DeclareSymbolFont{hebrewletters}{U}{rcjhbltx}{m}{n}

\newcommand{\rs}{\rm\scriptscriptstyle}

\usepackage{graphicx}
\usepackage{color}
\usepackage[usenames,dvipsnames]{xcolor}
\usepackage[colorlinks=true,linkcolor=Red,citecolor=Green,linktoc=page]{hyperref}
\usepackage{multirow}
\usepackage{float}
\usepackage{flushend}
\usepackage{balance}
\usepackage[varg]{txfonts}
\usepackage{ulem}
\usepackage{fancyhdr}

\DeclareMathSymbol{\lamed}{\mathord}{hebrewletters}{108}

\begin{document}
\title{Vogel-Fulcher-Tamman criticality of 3D superinsulators}
\author{	M.\,C.\,Diamantini}
\affiliation{NiPS Laboratory, INFN and Dipartimento di Fisica e Geologia, University of Perugia, via A. Pascoli, I-06100 Perugia, Italy}
\author{L. Gammaitoni}
\affiliation{NiPS Laboratory, INFN and Dipartimento di Fisica e Geologia, University of Perugia, via A. Pascoli, I-06100 Perugia, Italy}
\author{C.\,A.\,Trugenberger}
\affiliation{SwissScientific Technologies SA, rue du Rhone 59, CH-1204 Geneva, Switzerland}
\author{V.\,M.\,Vinokur}
\affiliation{Materials Science Division, Argonne National Laboratory, 9700 S. Cass Ave, Argonne, IL 60439, USA}

\begin{abstract}
It has been believed that the superinsulating state which is the low-temperature charge Berezinskii-Kosterlitz-Thouless (BKT) phase can exist only in two dimensions. We develop a general gauge description of the superinsulating state and the related deconfinement transition of Cooper pairs and predict the existence of the superinsulating state in three dimensions (3d).
We find that 3d superinsulators exhibit Vogel-Fulcher-Tammann (VFT) critical behavior at the phase transition. This is the 3d string analogue of the Berezinski-Kosterlitz-Thouless (BKT) criticality for logarithmically and linearly interacting point particles in 2d. Our results show that singular exponential scaling behaviors of the BKT type are generic for phase transitions associated with the condensation of topological excitations. 
\end{abstract}

\maketitle

\section{Introduction}

The superinsulating state, having infinite resistance at finite temperatures\,\cite{Diamantini1996,Doniach1998,vinokur2008superinsulator,vinokurAnnals,cbkt} is the state dual to superconductivity,  endowed with a finite temperature infinite conductance. 
Originally\,\cite{Diamantini1996,Doniach1998}, the emergence of superinsulation was attributed to logarithmic Coulomb interactions in two spatial dimensions (2d) arising from the dimensional reduction of the effective Coulomb interactions due to the divergence of the dielectric constant $\varepsilon$\,\cite{vinokur2008superinsulator, vinokurAnnals} near the superconductor-insulator transition (SIT)\,\cite{Efetov1980,Haviland1989,Paalanen1990,Fisher1990,Fisher1990-2,fazio,Goldman2010} in disordered superconducting films. A more recent approach\,\cite{dtv} based on the condensation of magnetic monopoles\,\cite{goddardolive}, however, derived superinsulation as a result of the linear confinement of Cooper pairs by electric strings\,\cite{polyakov_original,polyakov}, which are the S-dual of Abrikosov vortices in superconductors\,\cite{mandelstam, thooft, polyakov_original}. This offers a more general view of superinsulation as a phenomenon that is not specific to two dimensions but can also exist in 3d systems.

This immediately poses a question about the experimental effect that could serve as a hallmark of superinsulation and that could, at the same time, unequivocally discriminate between the 3d and 2d superinsulators, exposing the linear nature of the underlying confinement. In disordered superconducting films that host superinsulating state at the insulating side of the superconductor-insulator transition (SIT), it is the charge Berezinskii-Kosterlitz-Thouless (BKT) transition\,\cite{Berezinskii1970,Berezinskii1971,Kosterlitz1972,Kosterlitz1973} that marks the emergence of the superinsulating state\,\cite{vinokur2008superinsulator,vinokurAnnals} and which was detected experimentally\,\cite{cbkt} by the BKT critical behavior\,\cite{Kosterlitz1974}  of the conductance $G\propto\exp[-b/\sqrt{|T/T_{\rs{CBKT}}-1|}]$, where $T_{\rs{CBKT}}$ is the temperature of the charge BKT transition and $b$ is a constant of order unity. This suggests that it is the conductance critical behavior that provides the criterion for identifying the superinsulating state.

The BKT critical scaling of the conductance follows from the exponential critical scaling of the correlation length\,\cite{Kosterlitz1974}
\begin{equation}
\xi_{\pm} \propto \exp\left[\frac{b_{\pm}}{\sqrt{|T/T_{\mathrm c}-1|}}\right] \,,
\label{bkt}
\end{equation}
where $\pm$ subscript labels $T>T_{\mathrm c}$ and $T<T_{\mathrm c}$ regions respectively, and $\xi_{-}$ is interpreted as the maximum size of the bound charge-anticharge pair.
It is known\,\cite{yaffe} that, in two dimensions, both logarithmic and linear confinement lead to the BKT
critical scaling, so we use the notation $T_{\mathrm c}$ for either the BKT or deconfinement transition temperature. Our goal is now to reveal how the deconfinement scaling of Eq.\,(\ref{bkt}) evolves in 3d systems.
We show below that, in 3d, the critical behaviour of superinsulators is modified to the Vogel-Fulcher-Tammann (VFT) critical form\,\cite{vft} 
\begin{equation}
\xi_{\pm} \propto \exp\left(\frac{b^{\prime}_{\pm}}{|T/T_{\mathrm c}-1|}\right) \,.
\label{vft}
\end{equation}
This behaviour is characteristic of the one-dimensional confining strings in 3d, where the world-surface elements interact logarithmically as particles in 2d. The VFT scaling is typical of glassy systems and has recently been derived\,\cite{vasin} for the 3d XY model with quenched disorder. Here we show that it arises naturally, without assuming any disorder,  for confining strings and is thus a signature for 3d superinsulators. 

\section{Confining strings}
The electromagnetic effective action of a superinsulator is given by\,\cite{dtv} $S^{\rs SI}\propto\sum_{x,\mu,\nu}[1-\cos(2e\ell^2F_{\mu\nu})]$, where $\{x\}$ represent the sites a $d$-dimensional lattice, $\ell$ is the corresponding lattice spacing, $e$ is the electron charge, and $F_{\mu\nu}$ is the electromagnetic field strength. This is  Polyakov's compact QED action\,\cite{polyakov_original, polyakov}: hence the conclusion\,\cite{dtv} that, in superinsulators, Cooper pair dipoles are bound together  into neutral ``mesons" by Polyakov's
confining strings\,\cite{polyakov_confining}. 
These strings have an action which is induced by coupling their world-sheet elements to a massive Kalb-Ramond tensor gauge field\,\cite{kalb}. They can be explicitly derived for compact QED\,\cite{quevedo}, the induced electromagnetic action for the superinsulator\,\cite{dtv} and for Abelian-projected SU(2)\,\cite{antonov1, antonov2}. Their world-sheet formulation is thus in term of a non-local, long-range interaction between surface elements\,\cite{confstrings1}. Best suited to derive physical and geometric properties of these strings, however is the corresponding derivative expansion truncated to a certain level $n$\,\cite{confstrings2} (we use natural units $c=1$, $\hbar = 1$), 
\begin{eqnarray}
S &&= \int d^2 \xi \sqrt{g}  g^{ab}{\cal D}_a x_\mu  V_n ({\cal D}^2){\cal D}_b x_\mu \ ,
\nonumber \\
V_n ({\cal D}^2) &&= t \Lambda^2 + \sum_{k=1}^{2n} {c_k \over \Lambda^{2k-2} } ({\cal
D}^2)^k\ ,
\label{nmodel}
\end{eqnarray}
where ${\cal D}_a$ are the covariant derivatives with respect to the induced metric
$g_{ab} = \partial_a x_\mu \partial_b x_\mu$ on the world-sheet ${\bf x}(\xi_0,
\xi_1)$ embedded in $D=d+1$-dimensional Euclidean space-time and $g$ is the metric determinant. 
$V_n ({\cal D}^2)$ expresses the level-$n$ truncated derivative expansion of the non-local interaction on the world-sheet, with $\lambda$ being the fundamental ultraviolet (UV) cutoff mass scale. 
The first term in the bracket provides the bare surface tension $2t$. The numerical coefficients $c_k$ are alternating in the sign\,\cite{confstrings1} so that the stable truncation must end with an even $ k = 2n$. In particular, the second coefficient is the stiffness parameter accounting for the string rigidity. In confining strings it is actually {\it negative}. The string is stabilized by the last term in the truncation which generates a string tension $\propto \Lambda^2/c_{2n}$ taking control of the fluctuations where the orientational correlations die off and leads to long range correlations, thus avoiding the crumpling affecting most string models\,\cite{kleinert}. For example, in the simplest version with $n=1$, the third term in the derivative expansion, the string hyperfine structure, contains the square of the gradient of the extrinsic curvature matrices and it suppresses the
formation of spikes on the world-sheet. 

In a general model, the parameters $c_k$ are free: the only condition that must be imposed on them is the absence of both tachyons and ghosts in the theory. This requires that the Fourier transform $V_n\left( p^2 \right)$ has no zeros on the real $p^2$-axis. The polynomial $V_n \left( p^2 \right)$ thus has $n$ pairs of complex-conjugate zeros in the complex $p^2$-plane.
The associated mass scales represent the $n$ string resonances determining the string structure on finer and finer scales, the first resonance being the hyperfine structure. Increasing $n$ amounts thus to measuring the string on ever finer scales. 
 To simplify computations one can set all coefficients with odd $k$ to zero, $c_{2m+1} =0$ for $0\le m \le n-1$. This, however, is no drastic restriction since, as was shown in \cite{kleinert}, this is their value at the infrared-stable fixed point anyhow. Of course, when deriving the confining string from compact QED, all coefficients $c_k$ are fixed in terms of the only two dimensionless parameters available, $e/\Lambda^{2-D/2}$, with $e$ the QED coupling constant, and the monopole fugacity $z^2$. In particular\,\cite{confstrings1}
 \begin{equation}
 t={z^2\over 4(2\pi)^{D/2-1}} \tau^{D/2-2} K_{D/2-2} (\tau) \ ,
 \label{baretension}
 \end{equation}
 where $\tau = \Lambda^{2-D/2} z/e$. 
 
In the following we will consider the confining string model at finite temperatures in the large $D$ approximation. In\,\cite{confstrings2} it was shown that the high temperature limit behaviour of confining strings matches the expected high-temperature behaviour of large $N$ QCD\,\cite{polchinski}. Here we will, instead, concentrate on the critical behaviour at the deconfinement transition, where the renormalised string tension vanishes and strings become infinitely long on the cutoff scale. 

\section{Finite temperature behaviour}
Following\,\cite{kleinert2, david}, we introduce a Lagrange multiplier 
$\lambda^{ab}$  that forces the induced metric  
$\partial_a x_\mu \partial_b x_\mu$ to be equal to the intrinsic metric $g_{ab}$. The action becomes thus 
\begin{equation}
S\rightarrow S + \int d^2 \xi \sqrt{g} \left[\Lambda^2\lambda^{ab} (\partial_a x_\mu \partial_b x_\mu - g_{ab}
)  \right] \ . \label{cmodel} 
\end{equation}
We then parametrize the world-sheet in a Gauss map by $x_\mu (\xi) = (\xi_0, \xi_1,
\phi^i(\xi)),\ i = 2,...,D-2$. $\xi_0$ is taken as a periodic coordinate satisfying
$-\beta/2 \leq \xi_0 \leq \beta/2$, with $\beta = 1/T$ and $T$ the temperature, while 
$-R/2 \leq \xi_1 \leq R/2$, $R$ being the string length. Finally, the $\phi^i(\xi)$ describe the $D-2$ transverse
fluctuations. We will be looking for a saddle-point solution with a diagonal metric $g_{ab}
= {\rm diag}\ (\rho_0, \rho_1)$, and a Lagrange multiplier of the form
$\lambda^{ab} = {\rm diag}\ (\lambda_0/\rho_0, \lambda_1/\rho_1)$.
With this Ansatz the action becomes the combination of a tree-level contribution $S_0$ and a fluctuations contribution $S_1$,
\begin{eqnarray}
S &&= S_0 + S_1\nonumber \\
S_0 &&=  A_{\rm ext}\ \Lambda^2 \sqrt{\rho_0 \rho_1} \left[ \right. t \left( {\rho_0 + \rho_1
\over \rho_0 \rho_1} \right) + \lambda_0 \left( { 1 -\rho_0 \over \rho_0}
 \right) 
+ \lambda_1 \left( { 1 -\rho_1 \over \rho_1} \right) \left. \right] \nonumber \ ,\\
S_1&&= \int d^2 \xi \sqrt{g} \left[ g^{ab}\partial_a \phi^i  V_n ({\cal D}^2) 
\partial_b \phi^i + \Lambda^2 \lambda^{ab} \partial_a \phi^i \partial_b \phi^i \right]\ ,
\end{eqnarray}
with $\beta R = A_{\rm ext}$ the extrinsic area in coordinate space. 
Integrating over the transverse fluctuations in the limit $R \to
\infty$ we get
\begin{equation}
S_1 = {D - 2\over 2} R \sqrt{\rho_1} \sum_{l = - \infty}^{+ \infty} \int {d
p_1\over 2 \pi} \ln\left[ 
 (p_1^2 \lambda_1 + \omega_l^2 \lambda_0) \Lambda^2 + p^2 V_n(p^2)  \right] \ ,
\end{equation}
where $p^2 = p_1^2 + \omega_l^2$, and  $\omega_l = {2 \pi \over \beta \sqrt{\rho_0} } l$.

We will now focus on temperatures such that 
\begin{equation}
{c_{2n}\over \Lambda^{4n-2}} {1\over \beta^{4n}} \gg \Lambda^2 t + \sum_{ k = 1 }^{2n -1} {c_{k} \over \Lambda^{2k-2}}
{1\over \beta^{2k}}\ .
\label{valid}
\end{equation}
In this case the highest-order term in the derivative expansion dominates the one-loop term $S_1$
when $ l \neq 0$. This $l \neq 0$ contribution can be computed by using analytic regularization and analytic continuation of the expression $\sum_{n =1}^\infty n^{-z} = \zeta(z)$ for the Riemann zeta function, with $\zeta(-1) = -
1/12$, 
\begin{eqnarray}
&&{D - 2\over 2} R \sqrt{\rho_1} \sum_{l = - \infty}^{+ \infty} \int {d
p_1\over 2 \pi}\ {\rm ln}\ {c_{2n}\over \Lambda^{4n-2}}
\left( \omega_l^2 + p_1^2 \right)^{2n +1}\nonumber \\
&&= {D - 2\over 2}\sqrt{\rho_1 \over \rho_0} (2n + 1) 4 \pi {R\over \beta} 
\sum_{l = 1}^{+ \infty}\sqrt{l^2}\\
&&=  -{D - 2\over 2}\sqrt{\rho_1 \over \rho_0}{(2n + 1) \pi \over3}{ R\over \beta}\ .
\end{eqnarray}

The calculation of the $l=0$ contribution
\begin{equation}
S_1 = {D - 2\over 2} R \sqrt{\rho_1}  \int {dp_1\over 2 \pi} \ln \left( p_1^2 \bar V_n(p_1^2)\right) \ ,
\label{lzero}
\end{equation}
with
\begin{equation}
\bar V_n(p_1^2) = 
\left( \Lambda^2 (t + \lambda_1) +  \sum_{k =1}^{2n} 
{c_{k} \over \Lambda^{2k-2}} p_1^{2k}\right)   \ ,\label{newpot}
\end{equation}
requires a bit more care. Since we have chosen all $c_k=0$ for all odd $k$ and we have imposed the physical
requirement that the model is ghost- and tachyon-free,
all pairs of complex-conjugate zeros of $\bar V_n\left( p_1^2
\right)$ lie on the imaginary axis and we can represent\,\cite{kleinert} $\bar V_n
\left( p_1^2 \right)$ as 
\begin{equation}
{\Lambda^{4n-2}\over c_{2n}} \ \bar V_n\left( p_1^2
\right) = \prod_{k=1}^n \left( p_1^4 + \alpha_k^2 \Lambda^4 \right) \ ,
\label{poten}
\end{equation}
with purely numerical coefficients $\alpha_k$. Using again an analytic regularization and the analytic continuation of the Riemann zeta function we obtain 
\begin{eqnarray}
S_1 &&= {D - 2\over 2} R \sqrt{\rho_1} \sum_{k = 1}^{n} \int {d
p_1\over 2 \pi} \ln\left( p_1^4   + \alpha_k^2 \Lambda^4  \right) \nonumber \\
&&= {D - 2\over 2} R \sqrt{\rho_1} \sum_{k = 1}^{n} \int {d
p_1\over 2 \pi} 2 {\rm Re} \ln\left( p_1^4   + i \alpha_k \Lambda^2 \right) 
\nonumber \\
&&= {D - 2\over 2} R \sqrt{\rho_1} \sum_{k = 1}^{n} \Lambda \sqrt{2 \alpha_k} \ .
\end{eqnarray} 
Summing finally the $l=0$ and the $l \neq 0$ contribution we obtain the full action
\begin{equation}
S = S_0 + {D - 2\over 2} R \sqrt{\rho_1} \left[\sum_{k = 1}^{n} \Lambda 
\sqrt{2 \alpha_k} -
{(2n + 1) \pi \over 3 \sqrt{\rho_0}}{ 1\over \beta}\right] \ .
\label{finac}
\end{equation}

The coefficients in the representation (\ref{poten}) are not entirely free. Indeed, in order to match (\ref{newpot}), the 
$p_1$-independent term must satisfy
\begin{equation}
\prod_{k=1}^n \alpha_k^2 \Lambda^4 = {\Lambda^{4n}\over c_{2n}}(t + \lambda_1)\ .
\label{gamcon}
\end{equation}
For simplicity's sake we shall also assume that all $\alpha_k$ are equal, implying essentially that there is a unique resonance that determines the fine details of the string oscillations. Then (\ref{gamcon}) implies:
\begin{equation}
\alpha^2_k = ( t + \lambda_1)^{1/n} \alpha^2\ ,\ \ \ \alpha = \left({1\over c_{2n}}\right)^{1/2n} \ .
\label{coeff}
\end{equation}
Since the fluctuations contribution $S_1$ is proportional to $(D-2)$, it is forced onto its ground state in the large $D$ limit. In this limit the metric components $\rho_0$ and $\rho_1$ and the Lagrange multipliers $\lambda_0$ and $\lambda_1$ take on their classical values obtained by setting the respective derivatives of the total action to zero. This gives the four large-$D$ gap equations
\begin{eqnarray}
&&{ 1 -\rho_0 \over \rho_0} =  0 \ , \label{gap1}\\
&&{1 \over \rho_1} = 1 - {D - 2\over 2} { 1 \over 4 \beta \Lambda} 
\sqrt{2 \alpha} (\lambda_1 + t )^{1/4n -1} \ , \label{gap2} \\
&&\left[ {1\over 2}(t - \lambda_1) + {1 \over 2\rho_1}(\lambda_1 + t ) - t -
\lambda_0 \right] + {D - 2\over 2} { (2n +1)\pi \over 6 \beta^2 \Lambda^2} = 0 \ ,\label{gap3} \\
&&(t - \lambda_1) - {1 \over \rho_1}(\lambda_1 + t ) +
\nonumber \\
&&+ 
{D - 2\over 2} {1 \over  \beta \Lambda} \left[\sqrt{2 \alpha}\ n \left( \lambda_1 
+ t \right)^{1/4n} - {\pi (2n+1) \over 3 \beta \Lambda } \right] = 0 
\label{gap4} \ .
\end{eqnarray}
Inserting (\ref{gap4}) and (\ref{gap1}) into (\ref{finac}) and using $\rho_0 =1$ from
(\ref{gap1}) we obtain the action in the form 
\begin{equation}
S= A_{\rm ext}\  {\cal T} \ ,
\label{effac} 
\end{equation}
with ${\cal T} =  \Lambda ^2 2 (\lambda_1 + t )/\sqrt{\rho_1}={\cal T}_0/\sqrt{\rho_1}$  representing the renormalized string tension, expressed in terms of the zero-temperature renormalized string tension ${\cal T}_0$.  Eq. (\ref{gap2}) for the spatial metric can be reformulated in the limit $n \to \infty$ as 
\begin{equation}
{1\over \rho_1} = 1- {D-2 \over 2} T {\sqrt{2\alpha} \Lambda \over 4} {1\over (\lambda_1+t) \Lambda ^2} \ ,
\label{spatialmetric}
\end{equation}
From here we recognize that the renormalization of the string requires taking the simultaneous limits $(\lambda_1 + t) \to 0$, $\sqrt{2\alpha} \to 0$ and $\Lambda \to \infty$ so that $(\lambda_1 + t) \Lambda^2$ and $\sqrt{2\alpha} \Lambda$ are finite. In this case both the $\rho_1$ metric element and the renormalized string tension acquire finite values. The scale $\sqrt{2\alpha} \Lambda$ represents the renormalized mass $M$ of the string resonance that determines, together with ${\cal T}_0$ all physical properties of the string. In particular, the finite temperature deconfinement critical behaviour is obtained as the limit $(\lambda_1+t) \to 0$. In this limit the strings become infinitely long on the scale of the cutoff and the particles at their ends are liberated. The critical behaviour is embodied by the behaviour of the (dimensionless) correlation length $\xi = 1/\sqrt{\lambda_1 +t} $ near the critical temperature. 

\section{Critical behaviour}

In order to study the critical behaviour we derive the gap equation for $(\lambda_1 + t)$ alone, by substituting (\ref{gap2}) into (\ref{gap4}). This gives 
\begin{equation}
(\lambda_1 + t ) -  {D - 2\over 2} {4 n +1 \over 8 \beta \Lambda}\sqrt{2\alpha}
\left( \lambda_1 + t \right)^{1\over 4n} +\ {D - 2\over 2}{2 n +1 \over  6} 
{ \pi \over   \beta^2\Lambda^2 } - t = 0 \ .
\label{quart}
\end{equation}
For $(\lambda_1 + t) \ll 1$ the first term can be neglected with respect to the second for large $n \gg 1$, which gives
\begin{equation}
4n\sqrt{2\alpha} (\lambda_1 + t)^{1\over 4n} = n {8\pi\over 3} {T\over \Lambda} -{2\over D-2} {8t\Lambda \over T} \ .
\label{firstmod}
\end{equation}
Dividing by $\sqrt{2\alpha}$ and subtracting on both sides of the equation a term $4n$ we obtain
\begin{equation}
4n\left[ (\lambda_1 + t)^{1\over 4n} -1\right] = {8\pi n\over 3} {T\over M} \left[ 1- {3  \over (D-2)\pi} {\vartheta \over T^2}  -{3 \over 2\pi}  {M\over T} \right] \ ,
\label{inter1}
\end{equation} 
where $\vartheta = 2t\Lambda^2/n$ is the bare string tension divided by $n$. The expression in square brackets on the right hand side can be formulated as
\begin{equation}
\left[ 1- {3  \over (D-2)\pi} {\vartheta \over T^2}  -{3 \over 2\pi}  {M\over T} \right] = 
\left( 1-{T_{+}\over T} \right) \left( 1+ {T_{\_} \over T} \right) \ ,
\label{plusminus}
\end{equation}
where
\begin{equation}
T_{\pm} = {4\over (D-2)} {\vartheta \over M} {1\over \mp1 + \sqrt{1+ {16 \pi \over 3(D-2)} {\vartheta \over M^2}}} \ .
\label{solutions}
\end{equation}
From here we read off the critical deconfinement temperature as $T_c = T_{+}$. As expected it is determined by a combination of the two mass scales $\sqrt{\vartheta}$ and $M$ in the model. Expanding the left-hand side of (\ref{inter1}) around this critical temperature we get
\begin{equation}
4n\left[ (\lambda_1 + t)^{1\over 4n} -1\right] = {8\pi \over 3}  \left[ 1+ \left( {T_{\_} \over T} \right) \right]  \left( {-n\Delta T\over M} \right)  - O\left( n\Delta T^2 \right) \ ,
\label{inter2}
\end{equation}
where $\Delta T = T_c-T$. 
Taking the limit $n\to\infty$, one obtains on the left-hand side ${\rm ln}(\lambda_1+t)$. The right-hand side, however, requires more care. Clearly, we recognize immediately that increasing $n\to \infty$ drives the string to its critical point $\Delta T=0$. To establish in detail how this occurs, however, we must resort to the behaviour (\ref{baretension}) of the bare string tension. For the relevant regime of small $\tau$ (strong coupling and low monopole fugacity) the quantity $\vartheta$ appearing in the equations above is given by
\begin{eqnarray}
\vartheta_{D=3} &&= {ze\over 8} {\Lambda^{3\over 2} \over n} \ ,
\nonumber \\
\vartheta_{D=4} &&= {z^2\over 4\pi} {\rm ln}\left( {2e\over z}\right)  {\Lambda^2\over n} \ ,
\label{scalingzero}
\end{eqnarray}
The zero-temperature fixed point being given by $\ell = 1/\Lambda =0$, this show that $n$ must scale as $n\propto \ell^{-3/2}$ for $D=3$ and $n\propto \ell^{-2}$ for $D=4$ in the approach to a fixed point. Correspondingly, we have $n\propto (\Delta T)^{-3/2}$ for $D=3$ and $n\propto (\Delta T)^{-2}$ for $D=4$ when approaching the finite temperature critical point. Therefore the gap equation
\begin{equation}
{\rm ln}(\lambda_1 + t) = {8\pi \over 3}  \left[ 1+ \left( {T_{\_} \over T} \right) \right]  {\rm lim}_{n\to \infty} \left( {-n\Delta T\over M} \right) + \dots \ ,
\label{finalgap}
\end{equation}
leads directly to the critical scaling behaviours given by Eq.\,(\ref{bkt}) and Eq.\,(\ref{vft}) when approaching the deconfinement transition from below, 
reproducing thus the BKT criticality predicted in $d=2$ by the Svetitsky-Yaffe conjecture\,\cite{yaffe} and  the VFT criticality of the deconfinement transition in $d=3$. 

Remarkably, exactly this 3d-like signature has been 
recently observed\,\cite{ovadia} for the finite temperature insulating phase in InO disordered films, in which the thickness is much larger than the superconducting coherence length. While it seems premature to view this result as a conclusive evidence, yet one can view it as a possible indication of  linear confinement in 3d superinsulators. Another important corollary of our results is that the disorder strength in disordered superconducting films plays the role of a parameter tuning the strength of the Coulomb interactions but that disorder in itself is irrelevant for the nature of the various phases around the SIT. Finally, an important and deep implication of our findings is that, since the VFT behavior of Eq.\,(\ref{vft}) is recognized as heralding glassy behavior, our results suggest that, in 3d, topological defects endowed with long range interactions generate a glassy state without any quenched disorder. Note that the VFT behavior in 3d superinsulators arises as characteristic of the deconfinement transition of a strongly interacting gauge theory. The putative glassy behavior below this transition heralds the formation of a {\it quantum glass} arising due to the condensation and entanglement of extended string-like topological excitations.

\subsection{Acknowledgments}
M. C. D. thanks CERN, where she completed this work, for kind hospitality. The work at Argonne (V.M.V.) was supported by the U.S. Department of Energy, Office of Science, Materials Sciences and Engineering Division.




\begin{thebibliography}{10}



\bibitem{Diamantini1996}
M.\,C.\,Diamantini, P.\,Sodano, C.\,A.\,Trugenberger,
Gauge theories of Josephson junction arrays.
{Nuclear Physics} B \textbf{474}, 641 -- 677 (1996).
  
\bibitem{Doniach1998}
A.\,Kr\"{a}mer, S.\,Doniach,
Superinsulator phase of two-dimensional superconductors.
{Phys. Rev. Lett.} \textbf{81}, 3523 -- 3527 (1998).

\bibitem{vinokur2008superinsulator}
,V.\,M.\,Vinokur,\ \textit{et~al.} Superinsulator and quantum synchronization.
{Nature} \textbf{452}, 613 -- 615 (2008). 

\bibitem{vinokurAnnals} T.\,I.\,Baturina,  V.\,M.\,Vinokur, Superinsulator-superconductor duality in
two dimensions. {Ann. Phys.} {\bf 331} 236 -- 257 (2013). 

\bibitem{cbkt}
A.\,Mironov {\it et al.} Charge Berezinskii-Kosterlitz-Thouless transition in superconducting NbTiN films. {Scient. Rep.} {\bf 8} 4082 (2018). 

\bibitem{Efetov1980} 
   K.\,B.\,Efetov,  Phase transition in granulated superconductors. 
   {Sov. Phys. JETP} \textbf{51}, 1015 -- 1022 (1980).

      
\bibitem{Haviland1989}
D.\,Haviland, Y.\,Liu, A.\,Goldman, Onset of superconductivity in the two-dimensional limit. 
{Phys. Rev. Lett}. \textbf{62}, 2180 -- 2183 (1989). 

\bibitem{Paalanen1990}
A.\,Hebard, M.\,A.\,Paalanen, 
Magnetic-field-tuned superconductor-insulator transition
in two-dimensional films. 
{Phys. Rev. Lett}. \textbf{65}, 927 -- 930 (1990).

\bibitem{Fisher1990} 
       M.\,P.\,A.\,Fisher, G.\,Grinstein, S.\,M.\,Girvin,
 Presence of quantum diffusion in two dimensions: Universal resistance at the superconductor-insulator transition. 
 {Phys. Rev. Lett}. \textbf{64}, 587 -- 590 (1990).
        

\bibitem{Fisher1990-2} 
    M.\,P.\,A.\,Fisher, Quantum Phase Transitions in Disordered Two-Dimensional Superconductors. 
        {Phys. Rev. Lett.} \textbf{65}, 923 -- 926 (1990).
    
         
\bibitem{fazio} 
        R.\,Fazio, G.\,Sch\"on,
         Charge and Vortex Dynamics in Arrays of Tunnel Junctions. 
        {Physical Review B},
         \textbf{43}, {5307 -- 5320} ({1991}).
         
         
\bibitem{Goldman2010}
A.,\,M.\,Goldman,  Superconductor-Insulator Transitions.
{Int. J. Mod. Phys.} B\textbf{24}, 4081 -- 4101 (2010).

\bibitem{dtv}
M.\,C.\,Diamantini, C.\,A.\,Trugenberger, V.\,M.\,Vinokur, Confinement and asymptotic freedom with Cooper pairs, submitted for publication. 

\bibitem{goddardolive}
P.\,Goddard,  D.\,I.\,Olive,  Magnetic monopoles in gauge field theories. {Rep. Prog. Phys. } {\bf 41} 1357 (1978). 

\bibitem{polyakov_original}
A.\,M.\,Polyakov, Compact gauge fields and the infrared catastrophe. 
{Phys. Lett.} {\bf 59} 82-84 (1975). 

\bibitem{polyakov}
A.\,M.\,Polyakov, {Gauge Fields and Strings}. Harwood Academic Publisher, Chur (Switzerland) (1987). 


\bibitem{mandelstam}
S.\,Mandelstam, Vortices and quark confinement in non-Abelian gauge theories. { Phys. Rep.} {\bf 23} 245-249 (1976). 

\bibitem{thooft}
G.\,'t\,Hooft,  in {High Energy Physics.} Zichichi, A. ed., Editrice Compositori, Bologna (1976). 

\bibitem{Berezinskii1970} 
V.\,L.\,Berezinskii, Destruction of long-range order in one-dimensional and two-dimensional
systems having a continuous symmetry group I. Classical systems. 
{Sov. Phys.--JETP} \textbf{32} 493--500 (1970).

\bibitem{Berezinskii1971}%
V.\,L.\,Berezinskii, 
{Zh. Eksp. Theor. Fiz}. \textbf{61}, 1144 (1971). (\textit{Sov. Phys.-- JETP}, \textbf{34}, 610-616 (1971)).
 
\bibitem{Kosterlitz1972}
J.\,M.\,Kosterlitz, D.\,J.\,Thouless, Long range order and metastability in two dimensional solids and superfluids. (Application of dislocation theory). 
{Journal of Physics C: Solid State Physics,} \textbf{5}, L124 (1972).

\bibitem{Kosterlitz1973} 
\,J.\,M.\,Kosterlitz, D.\,J.\,Thouless, Ordering, metastability and phase transitions in two-dimensioal systems. 
{J. Phys. C: Solid State Phys.}  \textbf{6} 1181--1203 (1973).

\bibitem{Kosterlitz1974}
J.\,M.\,Kosterlitz, The critical properties of the two-dimensional xy model. 
{J. Phys. C: Solid State Phys}.\,\textbf{7}, 1046 (1974).

\bibitem{yaffe}
B.\,Svetitsky, L.\,G.\,Yaffe, Critical behavior at finite temperature confinement transitions. {Nucl. Phys.} {\bf B210} [FS6] 423-447 (1982). 

\bibitem{vft} 
P.\,W.\,Anderson,  {Lectures on amorphous systems}; Les Houches, Session XXXI, 1978, Balian, R. et al. eds., North Holland, Amsterdam (1978). 

\bibitem{vasin}
M.\,G.\,Vasin, V.\,N.\,Ryzhov, V.\,M.\,Vinokur, Berezinskii-Kosterlitz-Thouless and Vogel-Fulcher-Tammann criticality in XY model. 
{arXiv:1712.00757} (2017).


\bibitem{polyakov_confining}
A.\,Polyakov, Confining Strings. { Nucl. Phys. } {\bf B486} 23-33 (1997).

\bibitem{kalb}
M.\,Kalb, P.\,Ramond, Classical Direct Interstring Action. { Phys. Rev. } {\bf D9} 2273 2284 (1974). 

\bibitem{quevedo}
M.\,C.\,Diamantini, F.\,Quevedo, C.\,A.\,Trugenberger,  Confining Strings with Topological Term. 
{Phys. Lett.} {\bf B396} 115-121 (1997). 

\bibitem{antonov1} 
D.\,Antonov, Gluodynamics string as a low-energy limit of the universal confining string. {Phys. Lett. } {\bf B427} 274 (1998). 

\bibitem{antonov2}
D.\,Antonov, D.\,Ebert, String representation of field correlators in the SU(3) gluodynamics. {Phys. Lett.} {\it B444} 208 (1998). 

\bibitem{confstrings1}
M.\,C.\,Diamantini,  C.\,A.\,Trugenberger, Surfaces with long-range correlators from non-critical strings. {Phys. Lett.} {\bf B421} 196-202 (1998).

\bibitem{confstrings2}
M.\,C.\,Diamantini, C.\,A.\,Trugenberger, Confining strings at high temperature. {JHEP} 0204:032 (2002). 
 
\bibitem{kleinert}
M.\,C.\,Diamantini, H.\,Kleinert,  C.\,A.\,Trugenberger,  Strings with negative stiffness and hyperfine structure. { Phys. Rev. Lett.} {\bf 82} 267-270 (1999). 
 
\bibitem{polchinski}
J.\,Polchinski, Z.\,Yang, High-temperature partition function of the rigid string.  {Phys. Rev.} {\bf D46} 3667 (1992). 

\bibitem{kleinert2}
H.\,Kleinert, Spontaneous generation of string tension and quark potential. {Phys. Rev. Lett.} {\bf 58} 1915 (1987).

\bibitem{david} 
F.\,David, E.\,Guitter, E. Rigid random surfaces at large d. {\it Nucl. Phys.} {\bf B295} [FS21] 332-362 (1988). 
 
 \bibitem{ovadia} M.\,Ovadia,  {\it et al.} Evidence for  a finite-temperature insulator.  {Scientific Reports} {\bf 5}  13503 (2015). 

\end{thebibliography}
\end{document}